# Stabilizing Solution-Substrate Interaction of Perovskite Ink on PEDOT:PSS for Scalable Blade Coated Narrow Bandgap Perovskite Solar Modules by Gas Quenching

Severin Siegrist, Johnpaul K. Pious, Huagui Lai, Radha K. Kothandaraman, Jincheng Luo, Vitor Vlnieska, Ayodhya N. Tiwari, Fan Fu*

Laboratory for Thin Films and Photovoltaics

Empa - Swiss Federal Laboratories for Materials Science and Technology

Ueberlandstrasse 129, CH-8600 Duebendorf, Switzerland

Corresponding Author*

fan.fu@empa.ch

TOC Graphics

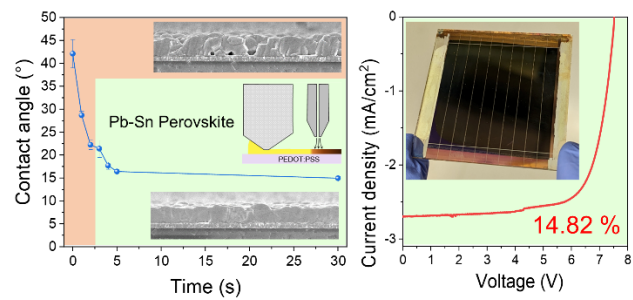

# Abstract

The development of scalable 1.25 eV mixed Pb-Sn perovskite solar modules by blade coating lags behind Pb-based perovskites due to limited understanding of solution-substrate interaction of the perovskite ink on PEDOT:PSS and subsequent gas quenching. To address this challenge, we systematically studied the wet film deposition and quenching process to better understand narrow bandgap perovskite film formation on PEDOT:PSS. We found, the wetting of Pb-Sn perovskite ink on PEDOT:PSS is highly unstable over relevant coating time scales, causing the contact angles to decrease rapidly from 42° to 16° within seconds. This instability leads to localized irregularities in the wet film, resulting in uneven solvent extraction and inhomogeneous nuclei density. As a result, rough perovskite films with voids at the buried interface are obtained. To overcome this problem, we developed a quasi-static wetting process by reducing the blade coating speed, thereby stabilizing the wetting behavior of Pb-Sn perovskite precursor ink on PEDOT:PSS. This optimized process facilitates the deposition of high-quality, void-free Pb-Sn perovskite films with uniform thickness over 8 cm of coating length using moderate (1.4 bar) $N_2$ quenching. We achieved 20 % efficient narrow bandgap perovskite solar cells and mini-modules with 15.8 % active area efficiency on 15.9 $cm^2$.

## Introduction

Scalable 1.25 eV narrow bandgap (NBG) perovskite solar cells and modules are required to achieve high performance 2-terminal all-perovskite tandem solar modules on large area (> 10 cm$^2$). So far, the focus of the perovskite research community has been on improving the power conversion efficiency (PCE) of NBG mixed Pb-Sn perovskite solar cells (PSCs). First, the notorious Sn$^{2+}$ oxidation is alleviated through mixing antioxidants [1-4] or reducing agents [5-12] into the perovskite precursor solution or applying a post deposition treatment [13-17]. Second, trap states and trap densities of the NBG perovskite absorber are reduced by regulating the fast and inhomogeneous crystallization rates of Sn- and Pb- based nuclei [16-23]. However, in all these studies, the NBG perovskite film is deposited by spin coating with anti-solvent treatment, which is not a scalable deposition method [24,25].

Replacing spin coating with scalable blade coating is a suitable approach to deposit NBG perovskite films on large area substrates. In addition, the anti-solvent treatment can no longer be applied and has to be substituted by gas quenching or vacuum quenching to induce supersaturation [22,26-28]. As a result, the wet film deposition and the quenching process significantly differ from the anti-solvent treated NBG absorbers by spin coating. So far, only five works report about fabricating NBG PSCs and achieving PCEs over 18 % [29-33]. For example, Tan's group published three works on all-perovskite tandem solar cells, using high pressure (> 4 bar) N$_2$ quenching for the NBG perovskite precursor solution, achieving efficiencies up to 21.4 % [29,32,33]. Dai et al. achieved an efficiency of 20.1 % by installing an in-line heater for using hot (100 °C) N$_2$ gas to quickly dry the blade coated NBG film during quenching [31]. Alternatively, Abdollahi et al. equipped their vacuum quenching system with an additional N$_2$ flow to facilitate solvent extraction from the blade coated NBG perovskite films [30]. Despite these developments on the quenching process, the efficiencies lag far behind the record values of 23.8 % obtained with spin coated NBG perovskite absorbers [17,34,35]. This shows it remains challenging to fabricate high quality NBG perovskite absorbers by scalable deposition and quenching methods. We relate this to a poor understanding of the solution-substrate interaction of the blade coated NBG perovskite ink onto PEDOT:PSS hole selective layer and the as well as to an ineffective solvent extraction by gas quenching with detrimental impact on the NBG film formation.

In this work, we reveal that the wettability of the NBG perovskite ink on PEODT:PSS is highly unstable with quickly decaying contact angles from 42 ° to 16 ° in 5 seconds. This induces localized wet film irregularities and inhomogeneous crystallization after quenching, resulting in rough films with voids at the buried interface. Thus, prolonging the solution-substrate interaction is key to obtain a high quality NBG perovskite film. We developed a quasi-static wetting process by reducing the blade coating speed to stabilize and enhance the wetting of Pb-Sn perovskite precursor ink on PEDOT:PSS. As a result, high quality NBG perovskite films can be deposited by blade coating with room-temperature gas quenching and moderate working gas pressures of 1.4 bar. These uniform perovskite films are void-free and show no thickness gradient over a coating length of 8 cm. We achieve NBG perovskite cells with an efficiency of 20 % and modules with an active area efficiency of 15.8 % on 15.9 cm$^2$.

# Results and discussion

A $MA_{0.3}FA_{0.7}Pb_{0.5}Sn_{0.5}I_3$ perovskite composition with DMF:DMSO of 9:1 (vol:vol) is used to deposit 1.25 eV narrow bandgap perovskite absorbers by $N_2$-assisted blade coating onto PEDOT:PSS substrates. **Figure 1a** illustrates the perovskite fabrication process, which consists of three steps, the wet film deposition, film quenching and film annealing. To obtain a high quality perovskite film, the quenching conditions of the $N_2$ quenching must be adjusted to the deposited wet film thickness. Effective film quenching to obtain micrometer thick NBG films is even more challenging. Thus, a thorough understanding of the nucleation and crystallization dynamics induced by $N_2$ quenching is required. In **Figure 1b**, we show the result of a design experiment to decouple the wet film deposition and the $N_2$ quenching step. In this experiment, the wet NBG perovskite film is blade coated over the entire substrate, while only the first half of the substrate is subjected to $N_2$ quenching. Then, the as-quenched film is segmented into five sections 1 to 5. Sections 1, 3 and 5 indicate a quenching regime according to their visual appearance, while sections 2 and 4 label the transition area.

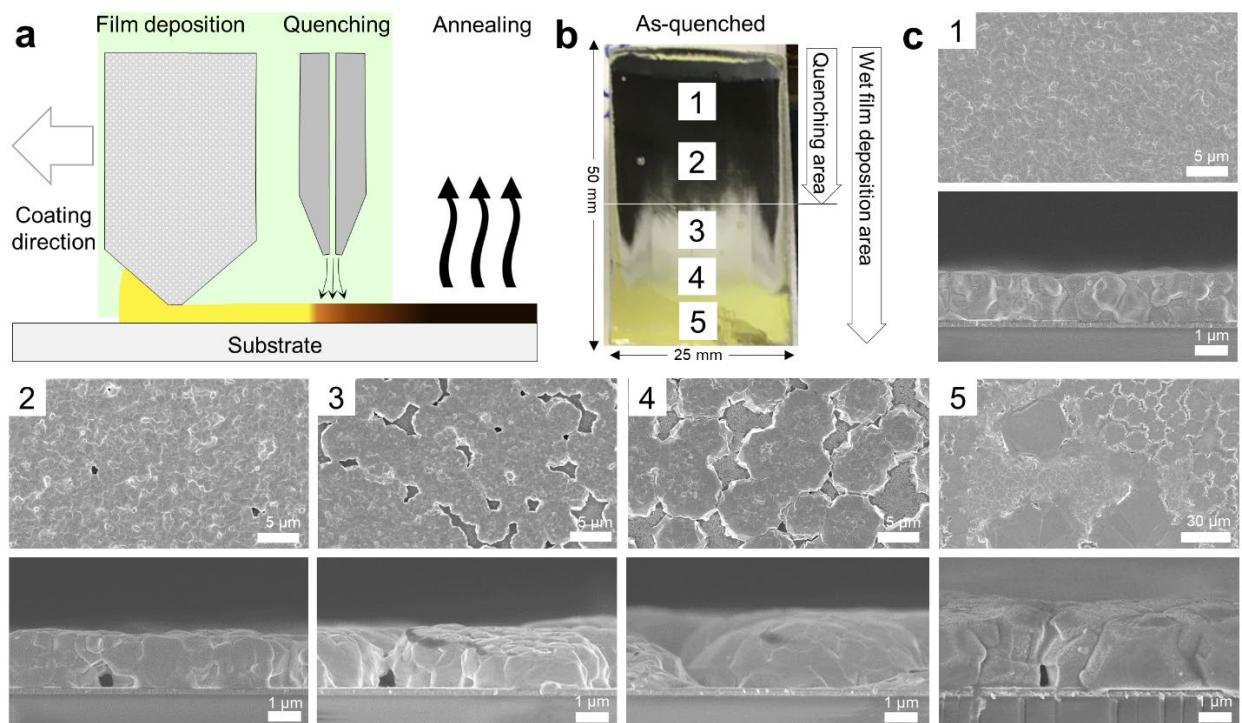

**Figure 1:** a) Schematics of the perovskite deposition by blade coating process, which is composed of the wet film deposition, gas quenching and thermal annealing. b): Photograph of the as-quenched NBG perovskite film. The precursor solution is deposited on the entire substrate, while the quenching is carried out only on the upper half. The film is zoned into different sections to reveal the three quenching regimes. c) SEM top and cross-section view images of the as-quenched NBG films at the corresponding sections from 1 to 5.

The first quenching regime, obtained in section 1, is characterized by a dry and opaque film. The solvent extraction in a relatively short time scale causes the concentration of the perovskite precursor solution to exceed the minimum supersaturation concentration $c_{min}^*$ by far, inducing a high nuclei density (**Figure S1**) [36,37]. During quenching, the crystallization has initiated. A compact, polycrystalline film with small grains and low surface roughness is formed as the crystal growth rate is smaller than the nucleation rate (**Figure 1c**) [38]. Upon thermal annealing, the small crystal grains coalesce and form larger crystals (**Figure S2b**). In section 2, the transition area from the first to the second quenching regime is characterized by a dry and opaque film after quenching. Pinholes have formed as result of a lower nuclei density, caused by an insufficient solvent extraction, which originates from the overlap of ceasing flow rate and adverse front streams, which are inherent to gas quenching (even when inclined angles or flow blockers are used) [39]. Refer to **Supplementary Note 1** for further details. Since the crystallization has already initiated during gas quenching, a solid shell has formed [39]. The film gets rougher and the overall film thickness increases due to mass conservation. Section 3 marks the second quenching regime. Here, the effect of the adverse front stream is disentangled from the main stream of the gas quenching. Even in the absence of the main stream, the adverse front stream is able to pre-dry the wet precursor film. The solvent extraction is slow and relatively weak (low mass transfer rate), resulting in an uncontrolled nucleation process, where the concentration of perovskite precursor solution exceeded the critical concentration $c_{min}^*$ marginally (**Figure S1**). The obtained film is dry, but transparent. The nucleation rate is lower than the crystal growth rate [37]. Hence, the nuclei density is low. Due to the low nuclei density, an increased number of pinholes emerge, which ends up in a dendritic, peninsula-like film with a rough film surface. The lateral growth of the individual perovskite crystal is no longer impinged by its neighbor. The annealed film has larger crystals, but pinholes too (**Figure S2b**). Section 4 is the transition area from the second to the third quenching regime. It is only marginally affected by the quenching of the front stream and marks the transition from peninsula-like to island-like (fragmented) films. Finally, section 5 indicates the third quenching regime, characterized by the complete absence of enforced quenching. The as-deposited film is transparent and remains wet, but gradually crystallizes after a few minutes. No additional nuclei are induced and the nuclei density in the wet film is determined by the presence of pre-nuclei in the wet precursor film (**Figure S1**). The lateral crystal growth is not restricted by adjacent grains, yielding large crystal clusters with increased film roughness and thickness after thermal annealing (**Figure S2b**). This island-like film is comparable to the observations by Dai et al. [31]. To conclude, we have shown that sufficient quenching is achieved with moderate working pressure of 1.4 bar for a 1 µm thick, pinhole-free NBG perovskite film on a reduced area of the substrate. This condition is used to study the wet film deposition step.

Next, we want to investigate the uniformity of blade coated NBG perovskite films and apply the condition of the first quenching regime over the entire substrate. We notice that the annealed film does not look uniform. **Figure 2a** shows the image of the film viewing from the substrate side. The film consists of two distinct regions. Region 1 appears black, while region 2 appears gray, occupying a major proportion of the coated area. To explain this different regions, we take a closer look at the blade coating process, which is illustrated in **Figure S3**. The blade coating process is composed of two steps: first, a certain volume of the perovskite precursor solution is dispensed onto the blade. The solution spreads across the width of the substrate and capillary flows drive the solution into the gap between the blade and the substrate. The excess of solution forms the advancing meniscus. The substrate area that is now covered by precursor solution is statically wetted and corresponds to region 1. In the second step, the blade coating is initiated by shearing the solution over the substrate with a constant speed. This remaining area is dynamically wetted and corresponds to region 2.

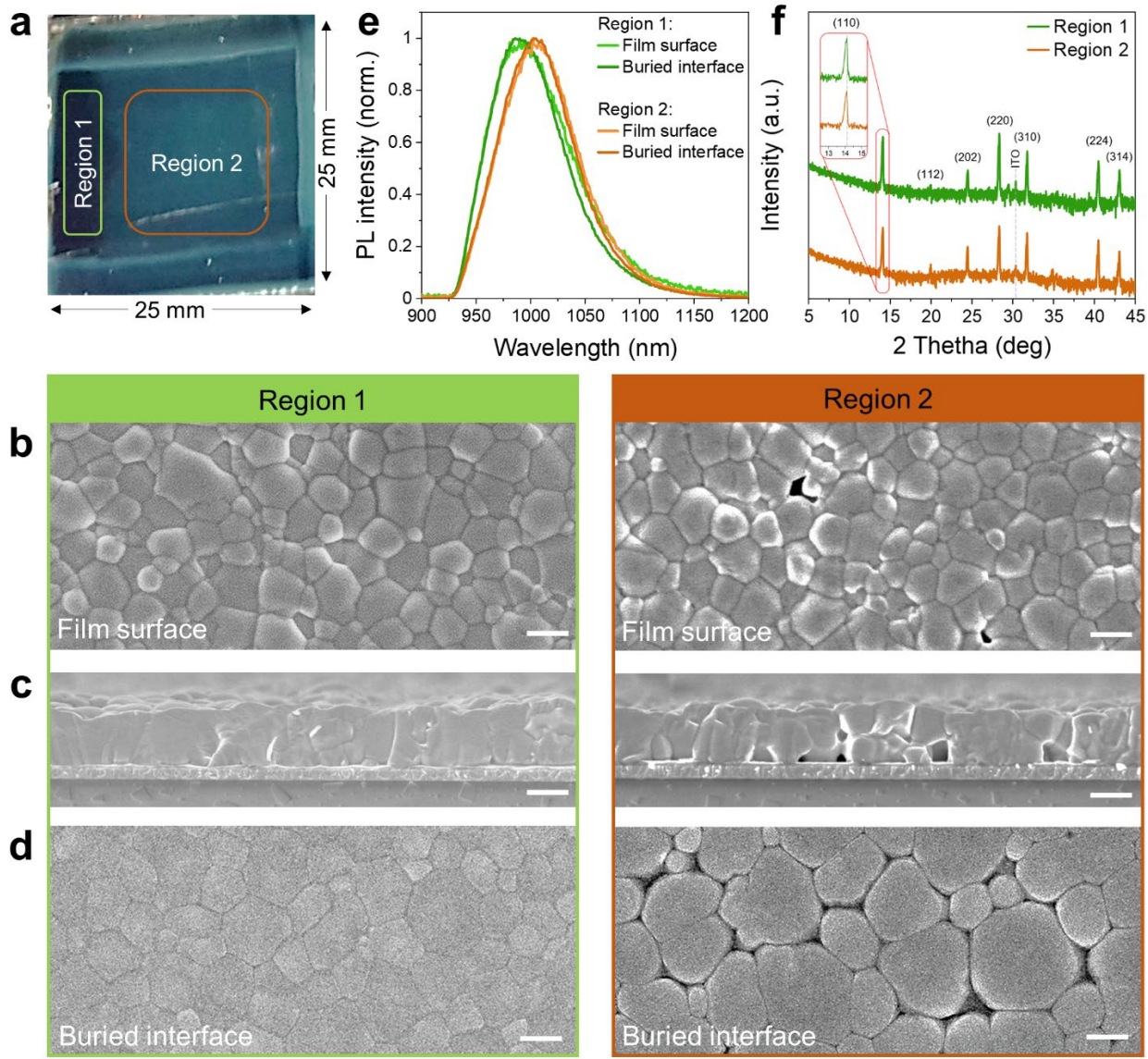

**Figure 2**: a) Photograph from substrate-view of blade coated NBG perovskite film. Two regions can be clearly differentiated. Region 1 marks the static wetting area, defined by the meniscus formation area prior to blade coating, while region 2 marks the dynamic wetting area, where the precursor solution is blade coated onto the substrate without resting. SEM images of region 1 and region 2 by b) top view, c) cross-section view and d) view on buried interface. The scale bar is 500 nm. e) Normalized photoluminescence spectra, probing the film surface and the buried interface of region 1 and region 2. f) XRD spectrum of region 1 and region 2.

In **Figure 2b-d**, SEM images of film surface, the film cross-section and the buried interface are presented. The top view SEM images show a void-free and compact perovskite film with grain size ranging from 200 nm to 700 nm for region 1, while region 2 shows a non-coherent perovskite film of similar grains size range, but with voids. AFM is used to quantify the surface roughness ($S_q$). Region 1 is smoother with a value of 26.3 nm, while region 2 returned 34.2 nm (**Figure S4**). The cross-sections show a comparable perovskite thickness of 600 nm for both regions. However, interfacial voids are present in region 2 with some voids even extending to the film surface of the perovskite film. These voids cause the higher surface roughness. To investigate the buried interface, the perovskite films are peeled-off from the substrate, following the method presented in the literature [39]. The buried interface of region 1 is composed of compact polycrystalline film. No voids are detected. The grain size range of the buried interface matches well to the top surface, yielding columnar grown crystals. In contrast, the buried interface of region 2 shows larger grain sizes compared to the film surface, leading to tapered grown crystals. In addition, these larger crystals have radiused edges, resulting in voids enclosing the individual grains. These voids at the buried interface are detrimental for long-term device stability, as presented in **Figure S5** and also reported in literature [39-41]. Steady-state photoluminescence (PL) spectra provided in **Figure 2e**, show a red-shifted PL peak for region 2 compared to region 1, indicating a bandgap narrowing by 12 meV. The PL peak position is not affected by the illumination direction. In **Figure S6a**, Tauc plots confirm an optical bandgap of 1.25 eV for region 1, and 1.24 eV for region 2. Bandgap narrowing can originate from compositional, structural and electronic factors [42,43]. As shown in **Figure S6b**, the Urbach energy increases from 15.2 meV in region 1 to 19.1 meV in region 2 [44,45]. This indicates an increased electronic disorder at the band onsets for region 2, but the Urbach energy is lower than the thermal energy $k_BT$ and a shift of the PL peak shift may not be expected [46,47]. Another factor may be a small compositional gradient in the Pb-to-Sn ratio from region 1 to region 2. However, XRD spectra of region 1 and region 2 with the major diffraction intensities at 14.09 °, 28.37 ° and 31.78 °, corresponding to (110), (220) and (310) crystal planes, are not shifted relatively, as presented in **Figure 2f**. Eperon et al. reported comparable results when increasing the Sn content from 50 % to 62.5 %, observing a PL redshift of 20 meV, while the crystal lattice parameters changed only marginally [48]. Thus, we hypothesize the bandgap narrowing may also originate from a gradient of the Pb-to-Sn ratio, while blade coating the perovskite precursor ink over the substrate. To sum up, static wetting is favored over dynamic wetting, leading to superior NBG perovskite quality with void-free interfaces and reduced electronic disorder.

Next, we elucidate the reason for the void formation at the buried interface during dynamic wetting by studying the interaction of the NBG perovskite precursor solution and the substrate. We measure the time-dependent contact angles of a 1.6 M $FA_{0.7}MA_{0.3}Pb_{0.5}Sn_{0.5}I_3$ NBG perovskite precursor solution on PEDOT:PSS. As reference substrates, we use UVO-treated ITO as hydrophilic and PTAA as hydrophobic representatives, respectively. The entire time series is provided **Figure S7** and the results are summarized in **Figure 3a**. The contact angle of the NBG solution on PEDOT:PSS is around 42 ° (0 s), then sharply drops to 30 ° (1 s) and stabilizing to 16 ° after 30 s. In contrast to PEDOT:PSS, the contact angle on PTAA and ITO do not alter at the same extent. For solution sheared coatings, these altering contact angles affect the characteristic length and thus the wet film thickness, resulting in wavy wet films of varying thickness [49,50].

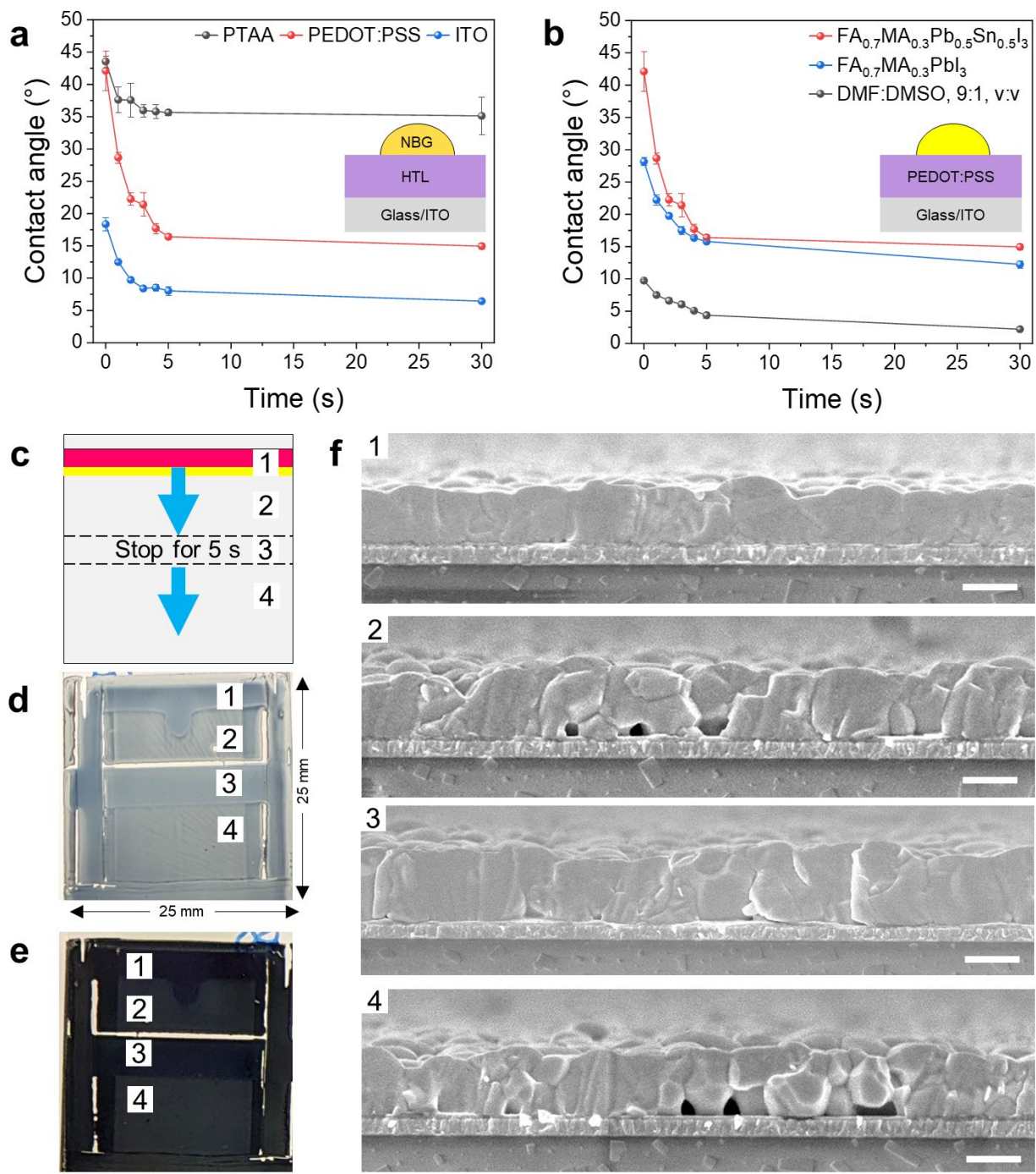

**Figure 3**: a) Time-dependent contact angle measurement of 1.6 M $FA_{0.7}MA_{0.3}Pb_{0.5}Sn_{0.5}I_3$ on different substrates. b) Time-dependent contact angle measurement of 1.6 M $FA_{0.7}MA_{0.3}Pb_{0.5}Sn_{0.5}I_3$, 1.6 M $FA_{0.7}MA_{0.3}PbI_3$ and the solvent system (DMF:DMSO, 9:1, vol:vol) on PEDOT:PSS. c) Top view of the experiment: the ink is sheared onto the substrate (PEDOT:PSS) at constant speed, then stopped for 5 s in the middle of the sample, before continuing with constant speed. d) Film side and e) substrate side view of the NBG perovskite film with labelled sections 1, 2, 3 and 4. f) Corresponding SEM cross-section view images of the four sections. The scale bar is 500 nm.

To check, whether this contact angle issue is only related to NBG perovskite solutions, we measured the contact angle for the pure lead (1.6 M $FA_{0.7}MA_{0.3}PbI_3$) and the solvent system of these perovskite solutions (DMF:DMSO, 9:1, by volume) on PEDOT:PSS. The summary is shown in **Figure 3b** and the whole time series in **Figure S8**. While the solvent system is almost perfectly wetting the substrate, the pure lead perovskite solution also shows decreasing contact angle, but the alteration is not as pronounced (relative decrease of 86.7 %) as for the mixed lead-tin perovskite solution (relative decrease of 165.6 %). To elucidate the reason for this difference, we characterize the rheological properties of these solutions. **Table S1** summarizes the surface tension, viscosity and density of the perovskite solutions and the solvent system. We observe upon alloying with tin, the surface tension of the mixed lead-tin perovskite solution is increased, yielding 59.48 mN/m, while for the pure lead perovskite solution, the surface tension yields 41.96 mN/m, which is similar to the solvent system's surface tension of 41.16 mN/m. Thus, the increased surface tension of the NBG perovskite solution results in increased contact angle on the PEDOT:PSS substrate, which is in accordance with Young's equation [51].

In the following, we perform two experiments to prove the enhanced NBG perovskite film quality by static wetting on PEDOT:PSS. By the experimental design, we exclude any unintentional impact of the gas quenching. In the first experiment, illustrated in **Figure S9a**, three perovskite precursor drops are dispensed onto the substrate, followed by a standard blade coating process. The resulting film shows three well-defined imprints of the drops (**Figure S9b, c**). These imprints appear black, while the remaining film area appears slightly grey and hazy area. In **Figure S9d**, SEM top view and cross-section view images are shown of the film area without the drops. Due to a smaller dispensed solution volume, the average film thickness decreases to 440 nm. The perovskite film is smooth and the buried interface is void-free. However, the perovskite crystals are small and some voids can be detected on the film side. Furthermore, white slender crystals appear on the film surface. We suspect, that $SnF_2$ has segregated out of the perovskite [52]. In contrast, these crystals cannot be detected on the film with the drop (**Figure S9e**). Additionally, the perovskite crystals have grown in size and the mean film thickness is increased to 620 nm due to the increased advancing meniscus height after merging with the pre-dispensed drop. No voids are detected. Thus, the sessile drops prolonged the interaction time of the perovskite solution with the substrate, improving the coating quality of the perovskite film. However, pre-wetting the substrate with perovskite precursor solution is not industrially feasible. Thus, we have to make use of the solution meniscus during blade coating, as proved in the next experiment. The second experiment is illustrated in **Figure 3c**. The perovskite precursor solution is sheared with constant speed onto the substrate. At halfway, the blade is stopped for 5 s, before it continues solution shearing with constant speed again. We divide the coating into two pairs. The first pair marks the static wetting, composed of section 1 (meniscus formation, start of blade coating) and 3 (stop for 5 s, interruption of blade coating). The second pair indicates the dynamic wetting area valid for section 2 and 4, where the solution is sheared at constant speed (5 mm/s). Viewing from the film side (**Figure 3d**), sections 1 and 3 appear blue, hinting at compact and smooth films, while sections 2 and 4 appear grey, indicating rough films. Notably, the longitudinal extent of section 3 matches well with the add-up length of the confinement (blade edge) and the length of the meniscus (**Figure S3b**). Viewing from the substrate side (**Figure 3e**), sections 1 and 3 appear black, indicating pinhole- and void-free films. In contrast, section 2 and 4 appear hazy, suggesting pinholes and voids in the film, which is also confirmed by the SEM cross-section view images in **Figure 3f**. To conclude these two experiments, we summarize our findings in **Figure S10**. In the case of static wetting, the transient contact angle is stabilized and a wet film of homogeneous thickness is deposited onto the substrate. The quenching conditions are sufficient to extract to solvents and induce a high nuclei density, which yields a coherent polycrystalline perovskite film without voids. In contrast, in the case of dynamic wetting, the contact angle of the solution and the substrate is highly fluctuating. As consequence, the wet film becomes wavy and the thickness is varying. The quenching conditions are not sufficient anymore to dry the film thoroughly. A low nuclei density is obtained, yielding a defective film with interfacial voids, caused after volatilization of the remaining solvents. Hence, static wetting is key to obtain a high quality perovskite film. But unlike spin coating, where static wetting is commonly done, and industrially feasible process requires a continuous deposition process, which means in the case of blade coating, we can only make use of a quasi-static wetting situation. In other terms, we have to prolong the solution-substrate interaction to enter the quasi-static wetting window of low and stable contact angles by reducing the coating speed (cf. **Figure 3b**).

To develop a continuous deposition process in the quasi static wetting condition, the coating speed is lowered step-wise from 5 to 1 mm/s. The reduction of the coating speed reduces the shear force accordingly, and thus the deposited wet film thickness. To compensate for the reduced thickness, we increase the dispensed solution volume accordingly. Increasing the solution volume increases the wetting area by the meniscus and the wet film thickness in turn. Moreover, as we have shown in our previous work, higher dispensed volumes reduce the difference of the advancing meniscus height (ΔH) along the coating direction [53]. Thus, we can expect a beneficial impact on the film uniformity with a minimized thickness gradient. **Figure S11** summarizes photographs of NBG perovskite films, deposited with reduced coating speed and increased volume. For 5 mm/s, the solution should be below 30 µL to avoid insufficient quenching (**Figure S11a**). The wet film thickness is decreased, when 3 mm/s is used and the film can be dried up to a volume that is smaller than 60 µL (**Figure S11b**). However, with 40 µL, the substrate side appears grey, indicating insufficient wetting time. Further decreasing the volume to 20 µL, the film gets smooth, but also very thin, shown by the increased pinhole density and the increased transparency of the film. Using 1 mm/s, which is the minimum speed of the blade coater, the film gets even more transparent for the case of 20 µL (**Figure S11c**). Pinholes appear in the perovskite film up to a volume of 40 µL, as highlighted in **Figure S11d**. With increased volume, the wet film thickness increases accordingly, and the pinhole formation is suppressed. Fully dried and void-free films can be obtained with 60 to 120 µL. SEM top view and cross-section view images of these three films are shown in **Figure S12**. Using 60 µL, the average film thickness is around 550 nm, with 90 µL it is 700 nm thick and with 120 µL, the film is around 1000 nm thick. Beyond 120 µL, the quenching is insufficient again, as the coordinating, non-volatile solvents are not completely extracted from the wet film (3rd quenching regime) [31,54].

To demonstrate the potential of blade coating using low coating speed and high volume, we compare two blade coating conditions in **Figure 4**. Condition 1 uses a regular coating speed of 5 mm/s and a dispensed solution volume of 15 µL (**Figure 4a**). In contrast, condition 2 operates in a quasi-static wetting window, using a slow coating speed of 1 mm/s and a dispensed solution volume of 80 µL (**Figure 4b**). Both conditions use a coating gap of 200 µm and the same quenching parameters (1.2 bar, 5 mm/s). Condition 1 yields a film that looks uniform only for the first 3 cm, before getting more transparent. The absorbance spectra of **Figure 4c** confirm this observation. The film thickness halves from 400 nm to 200 nm along the direction of the coating (**Figure 4e**). This is induced by a substantial decrease of the advancing meniscus height, as the solution is consumed from the meniscus during the coating, which results in a pronounced thickness gradient [55]. However, condition 2 yields a homogeneous film over the entire coating, confirmed by the sharp and coincide absorbance spectra (**Figure 4d**). Moreover, the film thickness is around 600 nm and is constant over the entire length, as shown in **Figure 4e**.

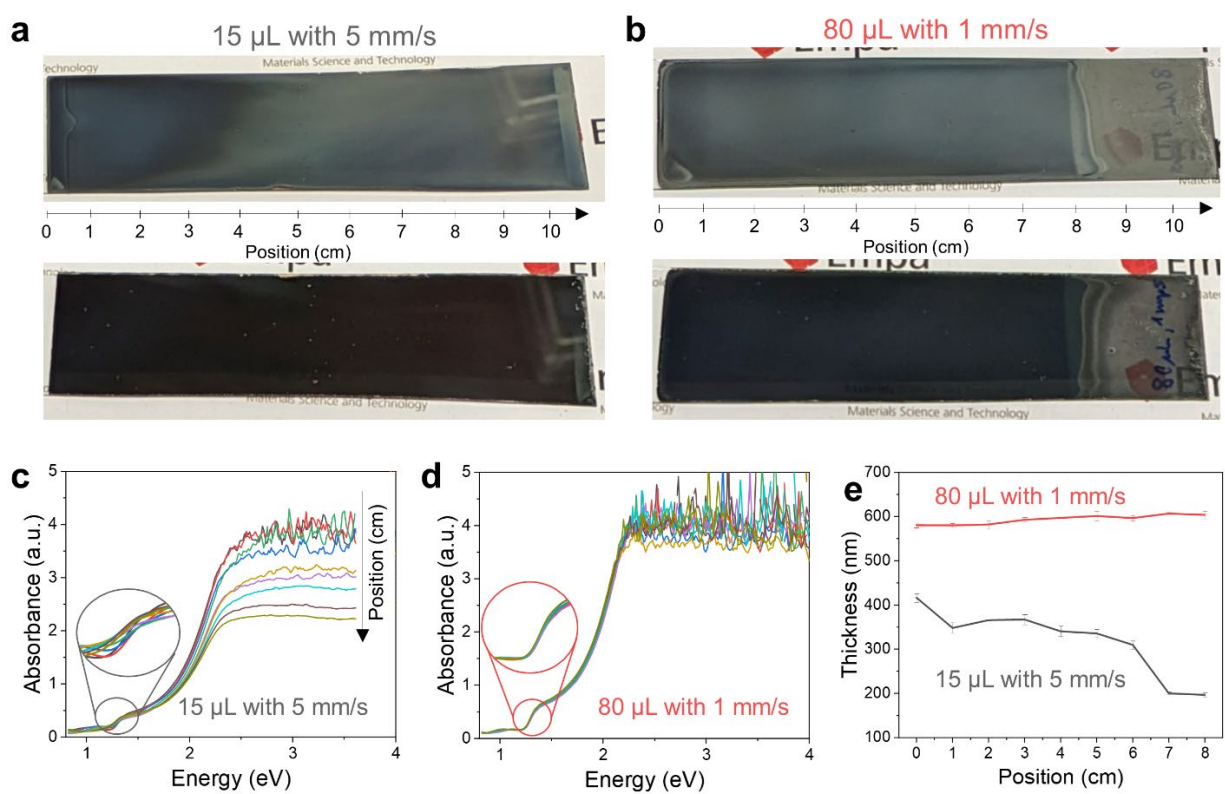

**Figure 4**: Photographs of film and substrate view on NBG perovskite films, based on a) fabrication condition 1 (coating speed of 5 mm/s with 15 µL volume) and on b) condition 2 (coating speed of 1 mm/s with 80 µL volume) with corresponding absorbance spectra shown in c, d) measured at 8 evenly spaced (1 cm) locations along the coating direction. e) Thickness measurements of the NBG perovskite film for both conditions taken along the coating direction.

As a last step, we fabricate p-i-n NBG PSCs composed of ITO/PEDOT:PSS/1.25 eV Perovskite/$C_{60}$/BCP/Cu (**Figure 5**). Using the optimum coating condition (1 mm/s, 80 µL) the PV performance and uniformity is greatly enhanced compared to condition 1 (**Figure S13**). We achieved a champion PCE of 19.2 % (**Figure 5b**). By blade coating $EDAI_2$ as post-deposition treatment, the performance exceeds 20 % with an open-circuit voltage ($V_{OC}$) of 853 mV and a fill factor (FF) of over 80 %, which is one of the highest reported efficiencies of blade coated NBG PSCs (**Table S2**) [17]. Maximum-power-point (MPP) tracking yields a stabilized efficiency of 19.5 mW/cm$^2$ (**Figure S14**). External quantum efficiency (EQE) gives an integrated $J_{SC}$ of 29 mA/cm$^2$ and an optical bandgap of 1.25 eV (**Figure 5c**).

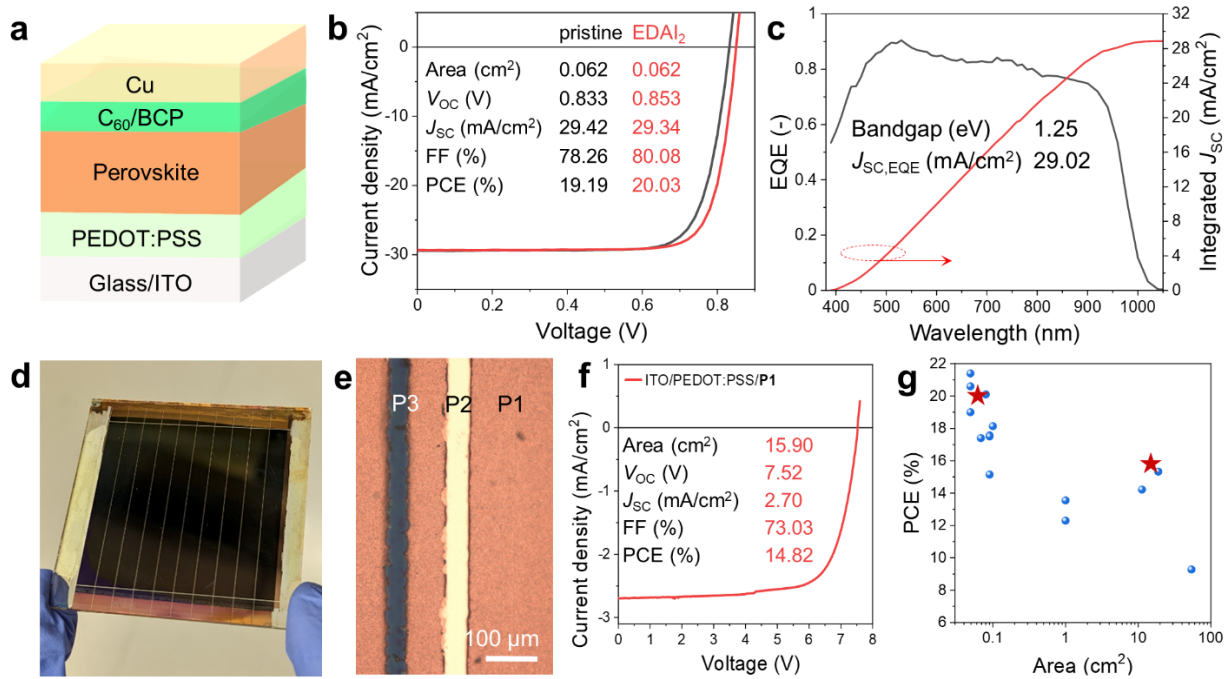

**Figure 5**: a) p-i-n layer stack of the NBG PSC. b) *J-V* comparison of pristine and EDAI$_2$-treated champion device. c) EQE measurement of the pristine champion device. d) Photograph of opaque NBG mini-module on 50 mm x 50 mm substrate. e) Image of laser scribes, P1, P2 and P3 accounting for the dead area. h) *J-V* scans of champion module with PEDOT:PSS/P1 scribing sequence. g) Comparison of active area efficiencies of blade coated NBG perovskite solar cells and modules.

To demonstrate the scalability of the NBG perovskite fabrication process, we fabricate single junction NBG perovskite mini-modules. Due to the relatively high conductivity of the organic p-type contact, we suspect that P1 can be shunted, resulting in a leakage current. Thus, we develop an alternative laser scribing sequence, illustrated in **Figure S15**, in which P1 is done after PEDOT:PSS deposition (denoted as ITO/PEDOT:PSS/P1), in contrast to the traditional scribing sequence, in which P1 is done prior to PEDOT:PSS deposition (denoted as ITO/P1/PEDOT:PSS). **Figure S16** summarized the comparison of these two sequences, by measuring the scribe profiles and the current-voltage characteristics. First, the quality of P1 on bare ITO is verified by the proper electrical isolation of the individual subcells. The traditional sequence ITO/P1/PEDOT:PSS shows that PEDOT:PSS can shunt the P1 trench, by partially filling it, as indicated by the reduced depth profile in **Figure S16d**. This can lead to a small leakage current in the sub-µA range, which potentially lowers the shunt resistance and thus the FF of the module. With the new laser scribing sequence, ITO/PEDOT:PSS/P1, the trench's depth increases again, but due to parasitic absorption by the PEODT:PSS layer, it is less deep than for the ITO/P1 case. The proposed scribing sequence yields proper insulation of the individual subcells. We then evaluate this sequence by fabricating mini-modules with an aperture area of over 15 cm$^2$ (**Figure 5d**). We achieve an aperture area efficiency of over 14.8 % (**Figure 5f**), with a stabilized power output of 14.1 mW/cm$^2$ (**Figure S17**). The mini-modules consist of 10 monolithically interconnected subcells, each measures 4 mm in width. The dead area is 250 µm, which yields a geometric fill factor of 93.75 % (**Figure 5e**). Thus, the active area efficiency is 15.8 %, which is among the highest reported efficiencies of opaque NBG modules with an active area over 10 cm$^2$ (**Figure 5g**) [29,55].

**Conclusion and outlook**

To conclude, we have systematically identified three quenching regimes of gas quenched NBG perovskite films to understand the interplay between quenching condition, nucleation and film morphology. To obtain compact, void-free perovskite films, a fast and strong solvent extraction is needed to obtain high nuclei density, as shown in the first quenching regime. It is imperative to clearly distinct the first quenching regime from the transition area, as voids and pinholes start to appear and the as-quenched film has already crystallized. In addition, we have further revealed an unsteady wettability of the NBG perovskite precursor solution on hydrophilic PEDOT:PSS, resulting in a pronounced decay of the contact angle at coating-relevant time scales. This altering contact angle in conjunction with high surface tension NBG perovskite precursor solution lead to localized variations of the wet film thickness and ultimately inducing voids in the perovskite film. We have developed a slow speed, high volume blade coating process to dynamically stabilize the wetting process of the NBG perovskite precursor solution on the PEDOT:PSS. Using this process, void-free NBG perovskite films can be fabricated. The resulting NBG PSCs deliver a PCE of over 20 %. Moreover, this process provides excellent uniformity by eliminating the thickness gradient along the blade coating direction and enabling 14.8 % efficient NBG modules with an aperture area of over 15 cm$^2$. We believe our findings and perovskite fabrication process will be useful for other labs and applied to various compositions and functional layers to drive down scaling losses of single junction and tandem solar modules by blade coating.

## ASSOCIATED CONTENT

*Supporting Information*. Experimental and characterization methods, schematics of quenching regimes, photographs and SEM images of annealed NBG perovskite films and their corresponding quenching regimes, schematics on static and dynamic wetting by blade coating, AFM, contact angle series, pre-wetting by droplets experiment, schematics on quenching and nucleation, photographs of NBG perovskite films with varied speeds and volume, SEM images of slow speed, high volume blade coating, PV performance heatmaps, MPP tracking of cells and modules, new P1 scribing process sequence, evaluation of P1 scribing sequence, supplementary note and tables of rheological parameters and PV performance of NBG PSCs, references.

*Notes*. The authors declare no competing financial interest.

## ACKNOWLEDGMENT

This project has received funding from the European Union's Horizon Europe research and innovation program under grant agreement No 101075605 as well as from the Swiss National Science Foundation (no. 200021_213073) and Swiss Federal Office of Energy (SI_502549-01).